\documentclass[twocolumn,trackchanges]{aastex701}

\begin{document}

\title{Expanding magnetic canopy structure and parasitic polarities in a quiet-Sun network element observed by {\sc Sunrise~iii}/SCIP}
\author[orcid=0000-0001-5699-2991]{Ryan J. Campbell}
\affiliation{Astrophysics Research Centre, Queen's University Belfast, Belfast, Northern Ireland, BT7 1NN, United Kingdom}
\email[show]{ryan.campbell@qub.ac.uk}

\author[orcid=0000-0001-5518-8782, gname=Carlos,sname=Quintero Noda]{Carlos Quintero Noda}
\affiliation{Instituto de Astrofísica de Canarias, Vía Láctea, s/n, E-38205 La Laguna, Spain}
\affiliation{Departamento de Astrof\'isica, Universidad de La Laguna, E-38206 La Laguna, Tenerife, Spain}
\email{carlos.quintero@iac.es}

\author[orcid=0000-0002-7725-6296,sname='Mathioudakis', gname=Mihalis]{Mihalis Mathioudakis}
\affiliation{Astrophysics Research Centre, Queen's University Belfast, Belfast, Northern Ireland, BT7 1NN, United Kingdom}
\email{m.mathioudakis@qub.ac.uk} 

\author[orcid=0000-0002-6210-9648]{Manuel Collados}
\affiliation{Instituto de Astrofísica de Canarias, Vía Láctea, s/n, E-38205 La Laguna, Spain}
\affiliation{Departamento de Astrof\'isica, Universidad de La Laguna, E-38206 La Laguna, Tenerife, Spain}
\email{manuel.collados@iac.es}

\author[orcid=0009-0009-9220-8231,sname='McFall', gname=Jack]{Jack McFall}
\affiliation{Astrophysics Research Centre, Queen's University Belfast, Belfast, Northern Ireland, BT7 1NN, United Kingdom}
\email{jmcfall01@qub.ac.uk} 

\author[orcid=0000-0002-9155-8039,sname='Jess', gname=David]{David B. Jess}
\affiliation{Astrophysics Research Centre, Queen's University Belfast, Belfast, Northern Ireland, BT7 1NN, United Kingdom}
\email{d.jess@qub.ac.uk} 

\author[orcid=0000-0002-3418-8449,sname='Solanki']{Sami~K.~Solanki} \affiliation{Max-Planck-Institut für Sonnensystemforschung, Justus-von-Liebig-Weg 3, 37077 Göttingen, Germany}\email{solanki@mps.mpg.de}

\author[orcid=0000-0003-1459-7074,sname='Lagg']{Andreas~Lagg} \affiliation{Max-Planck-Institut für Sonnensystemforschung, Justus-von-Liebig-Weg 3, 37077 Göttingen, Germany}\email{lagg@mps.mpg.de}
\author[orcid=0000-0002-9972-9840,sname='Gandorfer']{Achim~Gandorfer} \affiliation{Max-Planck-Institut für Sonnensystemforschung, Justus-von-Liebig-Weg 3, 37077 Göttingen, Germany}\email{gandorfer@mps.mpg.de}
\author[orcid=0000-0002-3387-026X,sname='del~Toro~Iniesta']{Jose~Carlos~del~Toro~Iniesta} \affiliation{Instituto de Astrofísica de Andalucía, CSIC, Glorieta de la Astronomía s/n, 18008 Granada, Spain}\affiliation{Spanish Space Solar Physics Consortium}\email{jti@iaa.es}
\author[orcid=0000-0002-5054-8782,sname='Katsukawa']{Yukio~Katsukawa} \affiliation{National Astronomical Observatory of Japan, 2-21-1 Osawa, Mitaka, Tokyo 181-8588, Japan}\affiliation{Department of Astronomy, The University of Tokyo, 7-3-1, Hongo, Bunkyo-ku, Tokyo 113-0033, Japan}\affiliation{Department of Astronomical Science, The Graduate University for Advanced Studies (SOKENDAI), 2-21-1 Osawa, Mitaka, Tokyo 1818588, Japan}\email{yukio.katsukawa@nao.ac.jp}
\author[orcid=0000-0002-0787-8954,sname='Bernasconi']{Pietro~Bernasconi} \affiliation{Johns Hopkins University Applied Physics Laboratory, 11100 Johns Hopkins Road, Laurel, Maryland, USA}\email{pietro.bernasconi@jhuapl.edu}
\author[sname='Berkefeld']{Thomas~Berkefeld} \affiliation{Institut für Sonnenphysik (KIS), Georges-Köhler-Allee 401a, 79110 Freiburg, Germany}\email{thomas.berkefeld@leibniz-kis.de}
\author[orcid=0009-0009-4425-599X,sname='Feller']{Alex~Feller} \affiliation{Max-Planck-Institut für Sonnensystemforschung, Justus-von-Liebig-Weg 3, 37077 Göttingen, Germany}\email{feller@mps.mpg.de}
\author[orcid=0000-0001-6317-4380,sname='Riethmüller']{Tino~L.~Riethmüller} \affiliation{Max-Planck-Institut für Sonnensystemforschung, Justus-von-Liebig-Weg 3, 37077 Göttingen, Germany}\email{riethmueller@mps.mpg.de}

\author[orcid=0000-0001-9228-3412,sname='Álvarez-Herrero']{Alberto~Álvarez-Herrero} \affiliation{Instituto Nacional de T\'ecnica Aeroespacial (INTA), Ctra. de Ajalvir, km. 4, E-28850 Torrejón de Ardoz, Spain}\affiliation{Spanish Space Solar Physics Consortium}\email{alvareza@inta.es}
\author[orcid=0000-0001-5616-2808,sname='Kubo']{Masahito~Kubo} \affiliation{National Astronomical Observatory of Japan, 2-21-1 Osawa, Mitaka, Tokyo 181-8588, Japan}\affiliation{Department of Astronomical Science, The Graduate University for Advanced Studies (SOKENDAI), 2-21-1 Osawa, Mitaka, Tokyo 181-8588, Japan}\email{masahito.kubo@nao.ac.jp}
\author[orcid=0000-0003-3490-6532,sname='Smitha']{H.~N.~Smitha} \affiliation{Max-Planck-Institut für Sonnensystemforschung, Justus-von-Liebig-Weg 3, 37077 Göttingen, Germany}\email{narayanamurthy@mps.mpg.de}
\author[orcid=0000-0001-8829-1938,sname='Orozco~Suárez']{David~Orozco~Suárez} \affiliation{Instituto de Astrofísica de Andalucía, CSIC, Glorieta de la Astronomía s/n, 18008 Granada, Spain}\affiliation{Spanish Space Solar Physics Consortium}\email{orozco@iaa.es}
\author[sname='Grauf']{Bianca~Grauf} \affiliation{Max-Planck-Institut für Sonnensystemforschung, Justus-von-Liebig-Weg 3, 37077 Göttingen, Germany}\email{grauf@mps.mpg.de}
\author[sname='Carpenter']{Michael~Carpenter} \affiliation{Johns Hopkins University Applied Physics Laboratory, 11100 Johns Hopkins Road, Laurel, Maryland, USA}\email{michael.carpenter@jhuapl.edu}
\author[sname='Bell']{Alexander~Bell} \affiliation{Institut für Sonnenphysik (KIS), Georges-Köhler-Allee 401a, 79110 Freiburg, Germany}\email{albe@leibniz-kis.de}
\author[orcid=0000-0001-7764-6895,sname='Martínez~Pillet']{Valentín~Martínez~Pillet} \affiliation{Instituto de Astrofísica de Canarias, Vía Láctea, s/n, E-38205 La Laguna, Spain}\affiliation{Spanish Space Solar Physics Consortium}\email{vmpillet@iac.es}

\author[orcid=0000-0002-7318-3536,sname='Bailén']{Francisco~Javier~Bailén} \affiliation{Instituto de Astrofísica de Andalucía, CSIC, Glorieta de la Astronomía s/n, 18008 Granada, Spain}\affiliation{Spanish Space Solar Physics Consortium}\email{fbailen@iaa.es}
\author[orcid=0000-0002-2055-441X,sname='Blanco~Rodríguez']{Julian~Blanco~Rodríguez} \affiliation{Universitat de Valencia Catedrático José Beltrán 2, E-46980 Paterna-Valencia, Spain}\affiliation{Spanish Space Solar Physics Consortium}\email{julian.blanco@uv.es}
\author[orcid=0000-0003-4319-2009,sname='Castellanos~Durán']{Juan~Sebastián~Castellanos~Durán} \affiliation{Max-Planck-Institut für Sonnensystemforschung, Justus-von-Liebig-Weg 3, 37077 Göttingen, Germany}\email{castellanos@mps.mpg.de}
\author[orcid=0009-0002-6808-5154,sname='Harnes']{Edvarda~Harnes} \affiliation{Max-Planck-Institut für Sonnensystemforschung, Justus-von-Liebig-Weg 3, 37077 Göttingen, Germany}\email{harnes@mps.mpg.de}
\author[orcid=0000-0001-6029-7529,sname='Hölken']{Johannes~Hölken} \affiliation{Max-Planck-Institut für Sonnensystemforschung, Justus-von-Liebig-Weg 3, 37077 Göttingen, Germany}\email{hoelken@mps.mpg.de}
\author[orcid=0000-0003-1409-1145,sname='Iglesias']{Francisco~A.~Iglesias} \affiliation{Max-Planck-Institut für Sonnensystemforschung, Justus-von-Liebig-Weg 3, 37077 Göttingen, Germany}\affiliation{Grupo de Estudios en Heliofísica de Mendoza, CONICET, Universidad de Mendoza, Boulogne sur Mer 683, 5500 Mendoza, Argentina}\email{iglesias@mps.mpg.de}
\author[orcid=0000-0002-4669-5376,sname='Ishikawa']{Ryohtaroh~T.~Ishikawa} \affiliation{National Institute for Fusion Science, 322-6 Oroshi-cho, Toki City 509-5292, Japan}\email{ishikawa.ryohtaro@nifs.ac.jp}
\author[orcid=0000-0001-7452-0656,sname='Kawabata']{Yusuke~Kawabata} \affiliation{National Astronomical Observatory of Japan, 2-21-1 Osawa, Mitaka, Tokyo 181-8588, Japan}\email{kawabata.yusuke@nao.ac.jp}
\author[orcid=0000-0002-1043-9944,sname='Matsumoto']{Takuma~Matsumoto} \affiliation{Centre for Integrated Data Science, Institute for Space-Earth Environmental Research, Nagoya University, Furocho, Chikusa-ku, Nagoya, Aichi 464-8601, Japan}\email{takuma.matsumoto@gmail.com}
\author[orcid=0000-0002-7044-6281,sname='Oba']{Takayoshi~Oba} \affiliation{Advanced Research Center for Space Science and Technology, Institute of Science and Engineering, Kanazawa University, Kakuma-machi, Kanazawa, Ishikawa 920-1192, Japan}\affiliation{Max-Planck-Institut für Sonnensystemforschung, Justus-von-Liebig-Weg 3, 37077 Göttingen, Germany}\email{oba@mps.mpg.de}
\author[orcid=0000-0003-0175-6232,sname='Siu-Tapia']{Azaymi~L.~Siu-Tapia} \affiliation{Instituto de Astrofísica de Andalucía, CSIC, Glorieta de la Astronomía s/n, 18008 Granada, Spain}\affiliation{Spanish Space Solar Physics Consortium}\email{siu@iaa.es}
\author[orcid=0000-0003-1483-4535,sname='Strecker']{Hanna~Strecker} \affiliation{Instituto de Astrofísica de Andalucía, CSIC, Glorieta de la Astronomía s/n, 18008 Granada, Spain}\affiliation{Spanish Space Solar Physics Consortium}\email{streckerh@iaa.es}
\author[orcid=0000-0003-1971-5551,sname='Vukadinović']{Dušan~Vukadinović} \affiliation{Institut für Physik, Universität Graz, Universitätsplatz 5, 8010 Graz, Austria}\affiliation{Max-Planck-Institut für Sonnensystemforschung, Justus-von-Liebig-Weg 3, 37077 Göttingen, Germany}\email{dusan.vukadinovic@uni-graz.at}


\author[orcid=0000-0001-5686-3081,sname='Hara']{Hirohisa~Hara} \affiliation{National Astronomical Observatory of Japan, 2-21-1 Osawa, Mitaka, Tokyo 181-8588, Japan}\email{hirohisa.hara@nao.ac.jp}
\author[orcid=0000-0003-4764-6856,sname='Shimizu']{Toshifumi~Shimizu} \affiliation{Department of Earth and Planetary Science, The University of Tokyo, 7-3-1, Hongo, Bunkyo-ku, Tokyo 113-0033, Japan}\affiliation{Institute of Space and Astronautical Science, Japan Aerospace Exploration Agency, 3-1-1, Yoshinodai, Chuo-ku, Sagamihara, Kanagawa 252-5210, Japan}\email{shimizu.toshifumi@isas.jaxa.jp}

\begin{abstract}

We present high-resolution multi-line spectropolarimetric observations of a quiet-Sun network element obtained with the {\sc Sunrise~iii} Chromospheric Infrared SpectroPolarimeter. The observations combine photospheric, upper-photospheric, and chromospheric diagnostics at a spatial resolution and polarimetric sensitivity that allow the transverse magnetic structure of the network boundary to be examined directly. We find that the strongest linear polarisation is concentrated in a narrow ridge around the edge of the magnetic element, co-spatial with enhanced transverse magnetic field inferred from multiline inversions. The magnetic azimuth exhibits a coherent, predominantly radial organisation around a more vertical core, consistent with an expanding magnetic canopy. An azimuth proxy derived directly from the observed Fe~\textsc{i} and K~\textsc{i} linear polarisation reproduces the same large-scale organisation, showing that this structure is encoded in the Stokes profiles rather than imposed by the inversion. Response functions and a MURaM-based forward-synthesis test indicate that the Fe~\textsc{i}~8468~\AA\ linear polarisation is sensitive to magnetic azimuth in the upper photosphere, with the closest proxy agreement occurring near $\log\tau\approx-3$. We find no evidence for strong azimuthal shear between the Fe- and K-sensitive diagnostics. At the network boundary, we also identify localised parasitic-polarity patches associated with complex, multi-lobed Stokes $V$ profiles, and one case in which the Stokes $V$ polarity reverses between photospheric Fe~\textsc{i} and chromospheric Ca~\textsc{ii} lines. These results demonstrate that quiet-Sun network boundaries contain organised upper-photospheric canopy fields together with small-scale mixed-polarity structure, providing new constraints on the three-dimensional magnetic structure of network elements.

\end{abstract}


\section{Introduction}

Quiet-Sun magnetic fields are organised into a hierarchy of structures spanning from the weak internetwork to strong, concentrated magnetic elements in the network. These network fields, located at the boundaries of supergranular cells, are thought to play an important role in coupling the photosphere to the chromosphere and above \citep{solanki2006}, but their resolved three-dimensional structure remains difficult to determine observationally \citep{rubio2019}.

In photospheric layers, small magnetic concentrations are commonly associated with bright structures in intergranular lanes and magnetic bright points, and are expected to evolve in geometry with height in a stratified atmosphere. Observational studies have provided evidence that such small-scale magnetic features expand and change their appearance between photospheric and higher-forming diagnostics, but detailed constraints on the vector magnetic field of resolved quiet-Sun network elements remain limited by the difficulty of measuring linear polarisation signals in particular \citep{Keys2019,Kuckein2019,marian2012}. For example, \citet{marian2012} inferred the expansion of a quiet-Sun network element indirectly from Stokes~$V$ asymmetries, but could not recover its magnetic azimuth and could only weakly constrain its inclination.

Simultaneous multiline spectropolarimetry offers a promising route to solving this problem. In particular, the spectral region sampled by the \textit{Sunrise Chromospheric Infrared spectroPolarimeter} (SCIP; \citealp{2026arXiv260317929K}) onboard the {\sc Sunrise~iii} balloon-borne solar observatory \citep{sunrise3} combines the photospheric Fe\,\textsc{i} $8514$~\AA\, Fe\,\textsc{i} $8515$~\AA, and Fe\,\textsc{i} 8468\,\AA\ lines with the chromospheric Ca\,\textsc{ii} 8498\,\AA\ and 8542\,\AA\ lines, providing sensitivity to both the photospheric and chromospheric magnetic field \citep{carlos2016}. The K~\textsc{i}~D$_1$~7699~\AA\ and K~\textsc{i}~D$_2$~7665~\AA\ lines provide an additional sensitivity to magnetic field strength in the mid-to-upper photosphere \citep{Klines2017}. The Fe\,\textsc{i} 8468\,\AA\ line is especially valuable because of its strong sensitivity to the magnetic field vector, including enhanced sensitivity to azimuth in the mid-to-upper photosphere compared with more commonly observed visible photospheric lines \citep{carlos2021}.

In this Letter, we use the unprecedented multiline capability of {\sc Sunrise iii}/SCIP to examine the magnetic structure of a quiet-Sun network element from the photosphere to the chromosphere. The key advance is the ability to combine high spatial resolution, sensitive polarimetry, and simultaneous Fe\,\textsc{i}, K\,\textsc{i}, and Ca\,\textsc{ii} diagnostics in a quiet-Sun network region. This allows us to test for the first time whether the transverse field structure expected from an expanding canopy is directly recoverable in the observed linear polarisation profiles, and to search for small-scale mixed-polarity structures at the network boundary. To our knowledge, however, no previous observation has \textit{directly} recovered the height-dependent vector structure of a  quiet-Sun network canopy, including its internal azimuthal organisation. For the first time, we are able to test whether that organisation remains coherent with height or develops azimuthal shear. We show that the network element contains a coherent transverse-field canopy surrounding a more vertical central region and that the inferred azimuthal organisation is recovered independently from the observed linear polarisation. Additionally, we find that localised parasitic-polarity patches occur at the boundary, including one case showing opposite Stokes~$V$ polarity between photospheric and chromospheric diagnostics.

\section{Observations}

On 11 July 2024 between 04:28:47--06:15:55 UT, observations of a quiet-Sun region at disk centre ($\mu = 1$) were obtained with the SCIP onboard the {\sc Sunrise~iii} balloon-borne solar observatory. The employed data were gathered as part of the observation with SUNRISE ID: {\sc 04$\_$QSUN} \citep{solanki2026}. The observations build on the heritage of earlier {\sc Sunrise} flights and instrumentation \citep{Barthol2011,Solanki2010,Solanki2017}. The flight took place under seeing-free conditions above the stratosphere. SCIP is a slit spectropolarimeter that scans in the direction perpendicular to the slit to build up two-dimensional maps of the solar surface. It operated in two spectral windows (SP1 and SP2), covering two spectral regions simultaneously. The spatial sampling along the slit and step size was $0{\,}.{\!\!}{\arcsec}094$, with a total of $621$ step positions, scanning a field of view of approximately $58{\arcsec}~\times~58{\arcsec}$.

The SP1 channel recorded a spectral region that includes the magnetically-sensitive photospheric Fe~I~$8468$~\AA\ line, as well as Fe~I~$8514$~\AA\ and Fe~I~$8515$~\AA, and the chromospheric Ca~II~$8542$~\AA\ and $8498$~\AA\ lines, while the SP2 channel covered the K~{\textsc{i}}~D$_{1}$ and K~{\textsc{i}}~D$_{2}$ lines. The mean cadence between consecutive slit-step positions was $10.4$~s. The standard deviation of the noise in continuum regions of Stokes~$Q$ and $V$ was $4.68\times10^{-4}\,I_\mathrm{c}$ and $4.16\times10^{-4}\,I_\mathrm{c}$, respectively, for SP1, where $I_\mathrm{c}$ is the spatially averaged continuum intensity. 

In this Letter we focus primarily on the analysis of the SP1 channel, specifically the Fe~I~$8468$~\AA, Fe~I~$8514$~\AA\, Fe~I~$8515$~\AA, Ca~II~$8498$~\AA, and Ca~II~$8542$~\AA\ lines, which together probe from the photosphere to the chromosphere. Where relevant, we incorporate the simultaneously observed K~I~D$_1$ and D$_2$ lines from the SP2 channel. We examine a network patch with significant linear and circular polarisation signals. The internetwork regions are pervaded by both linear and circular polarisation. In the Ca II lines, no linear polarisation is measured, but significant circular polarisation signals are detected in the network patches. Regions which are bright in the line core of the Fe I $8468$~\AA\ line are associated with the strongest magnetic concentrations. In the line core of the Ca II lines one can observe chromospheric canopy structures \citep{kubo2026}.

\begin{figure*}
\includegraphics[width=\linewidth]{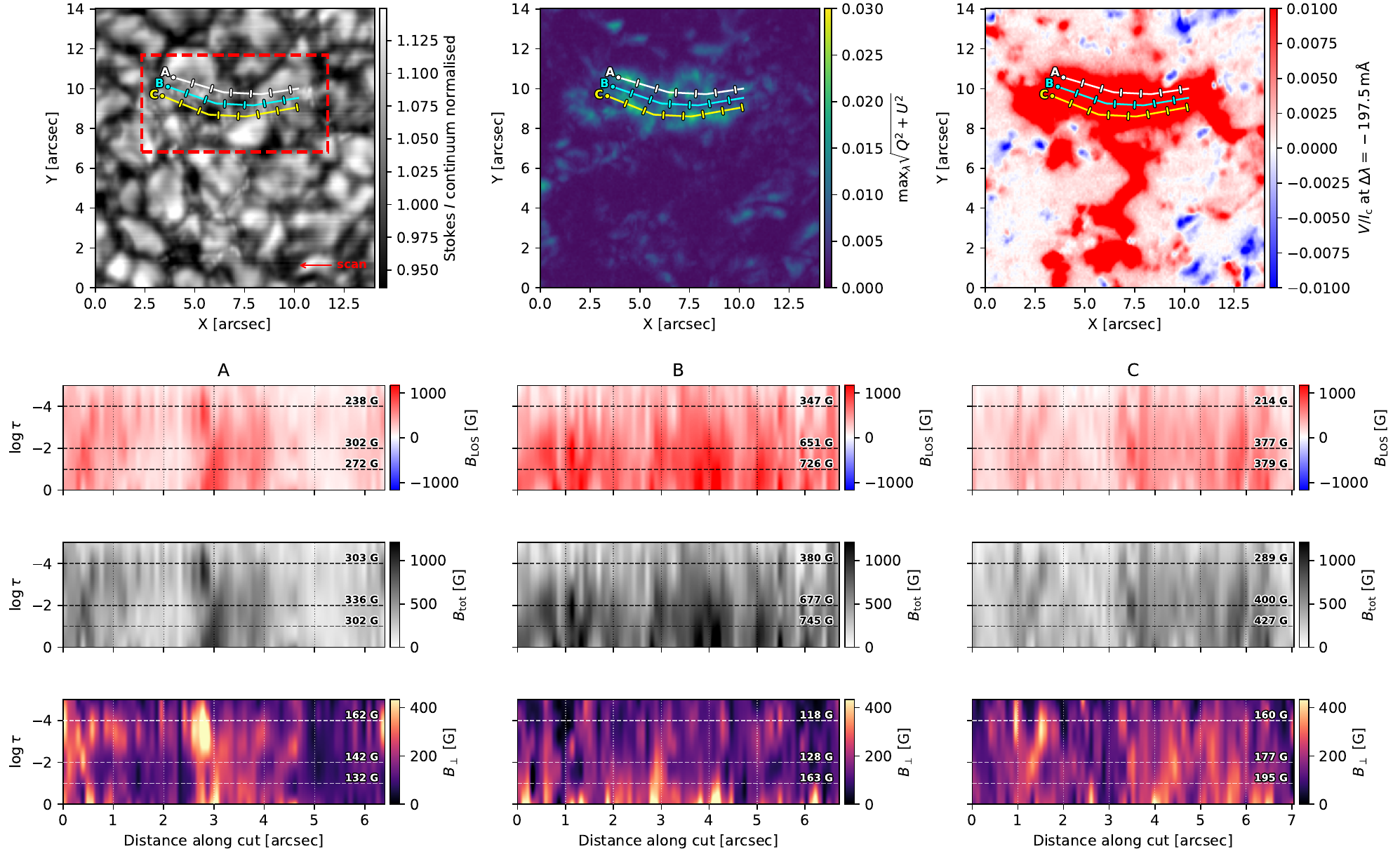}
\caption{Overview of the analysed network region and corresponding depth-dependent magnetic structure along three selected cuts. Top row: Continuum intensity (left), maximum linear polarisation signal $\max_{\lambda}\sqrt{Q^2 + U^2}$ (middle), and Stokes $V$ in the Fe~I~$8468$~\AA\ wing (right). Three spatial cuts (A, B, C) tracing the linear polarisation ridge are taken. Lower panels: Stratifications along cuts A–C as a function of distance and optical depth $\log \tau$. Rows show line-of-sight magnetic field $B_{\mathrm{LOS}}$, total field strength $B_{\mathrm{tot}}$, and transverse field component $B_{\perp}$, respectively. The dashed, red box highlights the region shown in Fig.~\ref{fig:azi}. All Stokes quantities are continuum normalised. The tick marks on the upper panels, and vertical lines on the lower panels, denote $1\arcsec$ increments. The median values at given optical depths are indicated on the bottom panels.
}
\label{fig:vector}
\end{figure*}

\section{Multiline inversions}

After standard data reduction \citep{solanki2026}, the full Stokes vector was inverted using the non-local thermodynamic equilibrium (NLTE) inversion code \textsc{DeSIRe} \citep{desire}. 
Photospheric Fe\,\textsc{i} lines were treated in LTE, while the K\,\textsc{i} and Ca\,\textsc{ii} lines were treated in NLTE. Elemental abundances were adopted from \citet{asplund}. Each pixel was modelled with a single magnetic atmosphere characterised by stratified parameters including temperature, magnetic field strength, inclination, azimuth, and line-of-sight (LOS) velocity. The microturbulent velocity, $v_{\mathrm{mic}}$, was included as a free parameter with one node, while the macroturbulent velocity, $v_{\mathrm{mac}}$ was also a free parameter primarily to account for the spectral point spread function. Inversions were performed across the full field of view ten times per pixel with varied initial conditions in $B$, $\gamma$, and $\phi$, with the minimum $\chi^2$ solution retained. The inversion was run through four cycles per pixel, with a final maximum number of nodes of $9$ in temperature and $6$ in the other parameters. Earlier cycles used fewer nodes. The magnetic filling factor was fixed to unity. We focus our analysis on inversion-derived properties of a network patch exhibiting strong and spatially coherent polarisation signals. 

\section{Magnetic structure of the network element}

\subsection{Height-dependent magnetic field properties}

The continuum intensity map (Fig.~\ref{fig:vector}, top left) shows a granular field with bright structures in intergranular lanes associated with magnetic flux concentrations. The corresponding Stokes $V$ map (Fig.~\ref{fig:vector}, top right) indicates that the region is primarily dominated by a single magnetic polarity, forming a spatially coherent patch. The maximum linear polarisation signal, defined as $\max_{\lambda}\sqrt{Q^2(\lambda)+U^2(\lambda)}$ across the Fe~I~$8468$~\AA\ line (Fig.~\ref{fig:vector}, top middle), reveals an elongated ridge co-spatial with the edge of the network element. This ridge or halo is not uniformly distributed across the magnetic patch, but instead traces a confined structure offset from the centre of the network element. Hereafter we refer to the central region devoid of significant linear polarisation as the ``core'' of the network element.

In order to examine the properties of the network structure, we describe the depth-dependent stratifications along three cuts labelled A–C. The cuts are intended to provide an illustrative example of the canopy structure, with the atmospheric quantities sampled along each path using bilinear interpolation. The exact placement of the cuts is not unique, but modest changes in their position or length do not alter the qualitative trends presented in this Letter. We show atmospheric parameters in this Section from the inversions that included the simultaneously observed K I lines from the SP2 channel. The results in Fig.~\ref{fig:vector} show that the line-of-sight magnetic field, $B_{\mathrm{LOS}}$, remains consistently of one sign across the structure at photospheric depths ($\log \tau \sim 0$ to $-2$), with peak values exceeding $\sim$kG in the core region (cut B). The total field strength $B_{\mathrm{tot}}$ has enhanced values concentrated in the central cut B and declines in cuts A and C. The $B_{\mathrm{tot}}$ and $B_{\mathrm{LOS}}$ both weaken in the chromosphere.

In contrast, the transverse component $B_{\perp}$ is enhanced along the same spatial locations as the linear polarisation ridge, with increased horizontal field strengths in cuts A and C. This indicates that the magnetic field becomes more inclined in these regions. The ratio of the median transverse to median total field strengths, $\langle B_\perp\rangle/\langle B_{\rm tot}\rangle$, provides a representation for the characteristic field inclination along each cut. At $\log\tau=-1$, this behaviour is reflected in median transverse-to-total field ratios of $\langle B_\perp\rangle/\langle B_{\rm tot}\rangle =0.44$ and $0.46$ for cuts A and C, respectively, compared to $0.22$ for the central cut B, with corresponding median inclinations of $27^\circ$ (A), $29^\circ$ (C), and $14^\circ$ (B). At $\log\tau=-2$, the contrast between the boundary and central cuts persists, with $\langle B_\perp\rangle/\langle B_{\rm tot}\rangle=0.42$ (A), $0.44$ (C), and $0.19$ (B), and median inclinations of $27^\circ$ (A), $29^\circ$ (C), and $12^\circ$ (B), respectively. This behaviour persists into higher layers. At $\log\tau=-4$, all three cuts exhibit substantially larger median transverse-to-total field ratios than in the photosphere, indicating that the magnetic field becomes significantly more horizontal with height. Nevertheless, the boundary cuts continue to show larger $\langle B_\perp\rangle/\langle B_{\rm tot}\rangle$ values ($\approx0.54$ (A), $0.55$ (C), and $0.31$ (B)).

\subsection{Transverse magnetic field structure}

\begin{figure*}
\includegraphics[width=.95\linewidth]{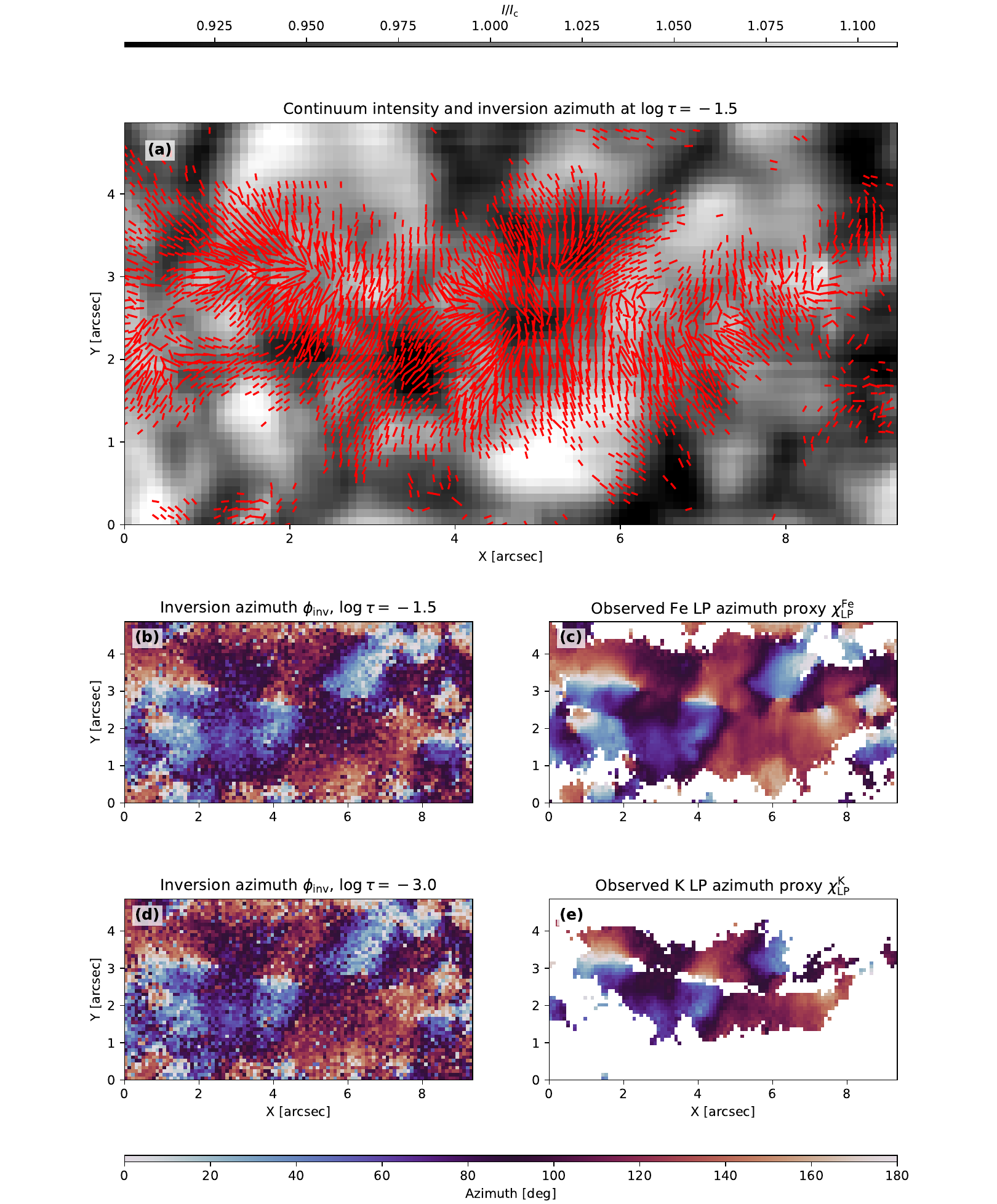}
\caption{Transverse magnetic field structure of the network element in the region highlighted by the dashed, red box in Fig.~\ref{fig:vector}. \textit{(a)} The background shows the continuum intensity, while red line segments indicate the corresponding magnetic azimuth, with lengths proportional to $B_{\perp}$. \textit{(b)} Magnetic azimuth $\phi_{\rm inv}$ inferred from the inversion at $\log \tau = -1.5$. \textit{(c)} Azimuth proxy $\chi_{\rm LP}$ derived directly from the observed linear polarisation of the Fe\,\textsc{i} $8468$~\AA\ line. Panels \textit{(d)} and \textit{(e)} show the same as \textit{(b)} and \textit{(c)} but for $\log \tau = -3.0$ and the K\,\textsc{i} $7699$~\AA\ line. In the lower panels, pixels are masked unless the amplitude of either Stokes Q or U exceeds $5\sigma_{QU}$ at one or more wavelengths. The close correspondence between the inversion and the proxy shows that the lateral variation in field orientation is already present in the observed Stokes profiles.}
\label{fig:azi}
\end{figure*}

The large amount of linear polarisation in the network region detected in Fe~I~$8468$~\AA~enables us to investigate the spatial variation of the inferred magnetic azimuth across it. In fact, there is significant linear polarisation detected in the K\,\textsc{i} $7699$~\AA\ line as well. For the avoidance of doubt, this analysis probes variations in field orientation in the plane of the solar surface. We conduct the analysis in this Section with the inversion that includes the K lines. The upper panel of Fig.~\ref{fig:azi} shows the continuum intensity in the region of strongest linear polarisation, overlaid with a quiver representation of the magnetic azimuth inferred from the inversion. The vector lengths scale with the transverse magnetic field strength, $B_{\perp}$, emphasising the locations where the transverse component is strongest. Enhanced transverse field is concentrated along the boundaries of the network element, while the interior remains comparatively vertical, forming a central core of weak $B_{\perp}$.

To assess whether the inferred azimuthal structure is present directly in the observations, we construct an azimuth proxy from the linear polarisation. For each pixel, we compute wavelength-integrated Stokes parameters over the Fe\,\textsc{i} $8468$~\AA\ line,
\begin{equation}
Q_{\rm int}(x,y) = \sum_{\lambda_1}^{\lambda_2} Q(\lambda, x, y), \quad
U_{\rm int}(x,y) = \sum_{\lambda_1}^{\lambda_2} U(\lambda, x, y),
\end{equation}
and define the corresponding azimuth proxy as
\begin{equation}
\chi_{\rm LP}(x,y) = \frac{1}{2}\,\mathrm{atan2}\big(U_{\rm int}(x,y),\, Q_{\rm int}(x,y)\big),
\end{equation}
which provides a formation-weighted estimate of the transverse field orientation, defined modulo $180^\circ$. \footnote{We note that in principle integrating across the Stokes $Q$ and $U$ profiles may incur cancellations, but we ultimately find the proxy is useful as a sanity-check on the inversion where the linear polarisation signals are strong.}  Fig.~\ref{fig:azi} shows the azimuth proxy derived directly from the observed linear polarisation. Without relying on any inversion assumptions, this proxy exhibits the same spatial organisation as the inversion-derived azimuth. The magnetic azimuths exhibit a coherent radial pattern centred on the most vertical region of the network element. Moving away from this central core, the transverse field orientation varies systematically around the structure, consistent with an expanding magnetic canopy. At the periphery of the network element, neighbouring regions often exhibit distinct azimuthal orientations, producing measurable lateral variations in the transverse field direction along the canopy boundary.

Panels \textit{(d)} and \textit{(e)} repeat this comparison using the K\,\textsc{i} D$_1$ 7699\,\AA\ line. The K-proxy preserves the same large-scale azimuthal organisation seen in Fe, and we find no evidence for a strong additional azimuthal shear between the Fe- and K-sensitive diagnostics within this network element; the azimuths from the two lines typically agree within $5-10^\circ$ of each other. This should not be interpreted as assigning either proxy to a single optical-depth layer; rather, both proxies represent contribution-weighted measurements of the emergent linear polarisation. 

\subsection{Comparison with MURaM simulations}\label{sect:MURaM}

\begin{figure*}
    \includegraphics[width=.95\textwidth]{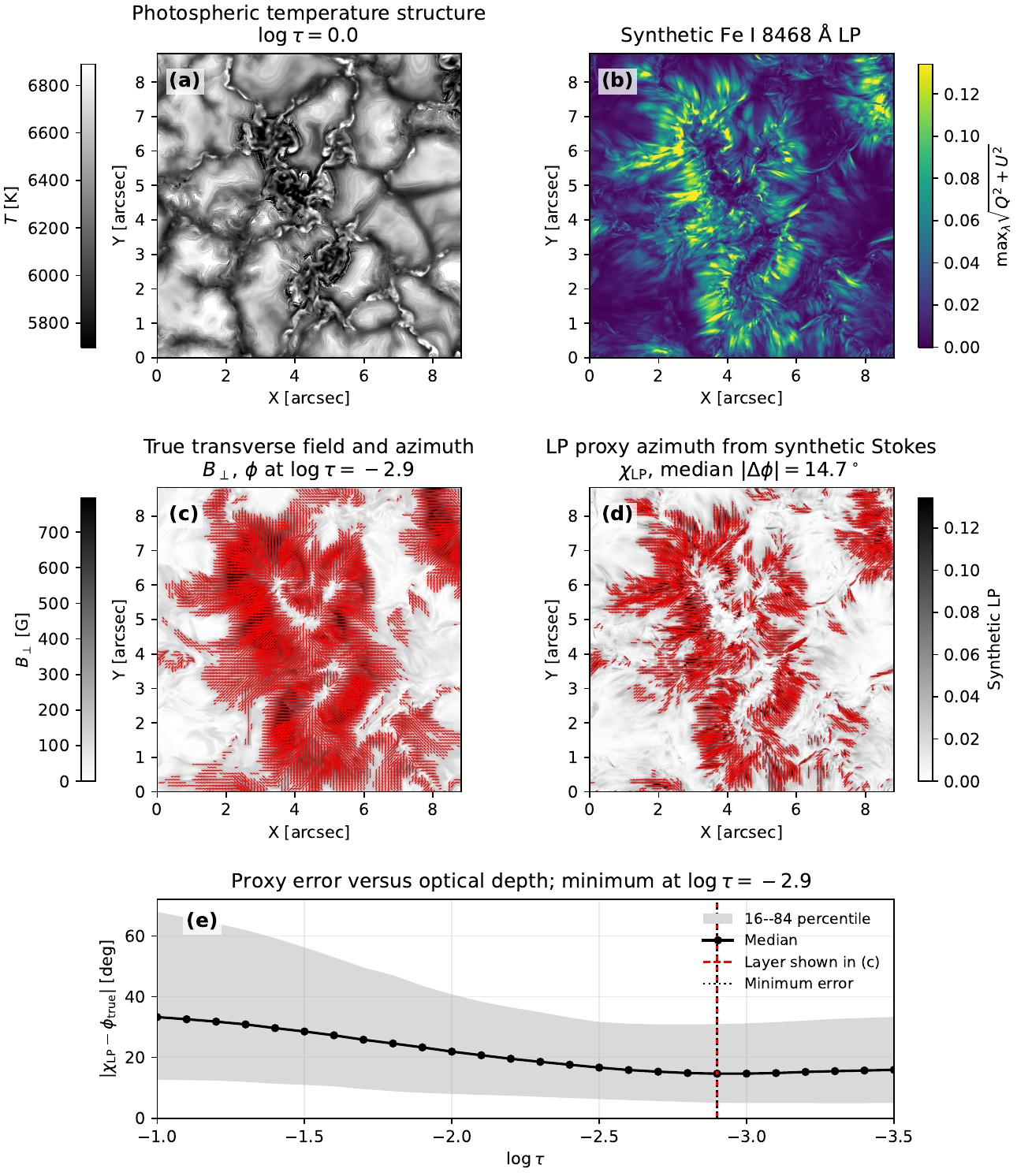}
    \caption{Illustrative magnetic structure of a quiet-Sun network element from a MURaM simulation. \textit{(a)} Temperature at $\log\tau=0$, showing a cool, concentrated magnetic core embedded within the granular photosphere. \textit{(b)} Maximum linear polarisation signal, $\max_{\lambda}\sqrt{Q^2+U^2}$, computed over the synthetic Fe~I~8468~\AA\ spectral window. The strongest linear polarisation occurs preferentially around the boundary of the magnetic concentration rather than within the central core. \textit{(c)} Transverse magnetic field strength $B_\perp$ at $\log\tau=-2$, with red lines indicating the magnetic azimuth directions. \textit{(d)} Azimuth proxy derived from the synthesised Stokes $Q$ and $U$ profiles using the same methodology applied to the SCIP observations, overlaid on the synthetic linear polarisation map. Red lines indicate the inferred azimuth orientation.  (e) Median absolute difference between the azimuth proxy, $\chi_{\rm LP}$, and the ground-truth magnetic azimuth in the MURaM simulation as a function of optical depth. The shaded region denotes the 16th–84th percentile range. The minimum error occurs near $\log\tau\approx-2.9$, indicating that the wavelength-integrated Fe~I~$8468$~$\mathrm{\AA}$ line is most sensitive to the magnetic azimuth at upper-photospheric heights.}\label{fig:muram}
\end{figure*}

To qualitatively place the inferred magnetic structure in a broader physical context, we examined a representative quiet-Sun network element from a radiative magnetohydrodynamic (rMHD) simulation produced with MURaM \citep{rempel}. The illustrative simulation shown in Fig.~\ref{fig:muram} was selected from the SPIN4D MURaM archive \citep{SPIN4D}, specifically from the \texttt{SPIN4D\_SSD\_Large} case. This simulation spans a larger horizontal domain ($50\times50$~Mm) than the other released cases while retaining the same high spatial resolution ($16$~km in the horizontal direction), allowing extended quiet-Sun network structures to develop self-consistently within realistic magnetoconvection. The selected network element was taken from a quadrant of the simulation containing relatively strong magnetic flux concentrations. The simulation therefore provides a physically plausible example of a mature quiet-Sun network concentration exhibiting organised transverse magnetic structure and spatially coherent azimuthal variation. While the comparison presented here is qualitative, the MURaM atmosphere offers a realistic magnetohydrodynamic context for interpreting the inferred morphology of the observed network element.

Figure~\ref{fig:muram} shows the photospheric temperature structure, the synthetic linear polarisation signal computed over the Fe~I~8468~\AA\ spectral window, and the transverse magnetic field and azimuth at $\log\tau=-2$. The Fe~I~8468~\AA\ Stokes profiles were forward synthesised from the MURaM atmosphere using DeSIRe in synthesis mode, allowing the spatial distribution of the synthetic linear polarisation signatures to be compared qualitatively with the SCIP observations. The simulated network element exhibits several qualitative similarities to the SCIP observations. The magnetic structure is anchored in a cool, vertically dominated central core with comparatively weak transverse field and linear polarisation signal, while enhanced linear polarisation and transverse magnetic field occur preferentially around the boundary of the flux concentration. The inferred azimuthal structure is also spatially radial, with coherent lateral variation in field orientation around the network element.

An additional comparison is provided in the fourth panel of Figure~\ref{fig:muram}, which shows the azimuth proxy derived directly from the synthesised Stokes $Q$ and $U$ signals using the same methodology applied to the SCIP observations. The proxy generally reproduces the large-scale organisation of the underlying magnetic azimuth, including the radial structure around the flux concentration and the coherent lateral variations along its boundary. A pixel-by-pixel comparison yields a median absolute difference of $14.7^\circ$ between the proxy and the ground truth azimuth at $\log\tau=-2.9$. Although local discrepancies are present, the agreement demonstrates that the proxy retains substantial information on the underlying magnetic field orientation and is capable of recovering the dominant azimuthal morphology present in the simulation. The comparison also highlights the limitations of any azimuth diagnostic derived directly from observed linear polarisation. Fine-scale variations visible in the simulation are partially suppressed or lost when the azimuth is inferred from the emergent Stokes signals. Degradation arising from finite spatial resolution, instrumental noise, and line-of-sight averaging would further reduce the fidelity with which the underlying azimuthal structure can be recovered. Most importantly, this demonstrates the Fe~I~8468~\AA\ line has significant sensitivity to the transverse magnetic field orientation in the upper photosphere. Appendix~\ref{sect:RFs} shows this is consistent with the response functions.

\subsection{Small-scale parasitic magnetism}

\begin{figure*}
    \includegraphics[width=\linewidth]{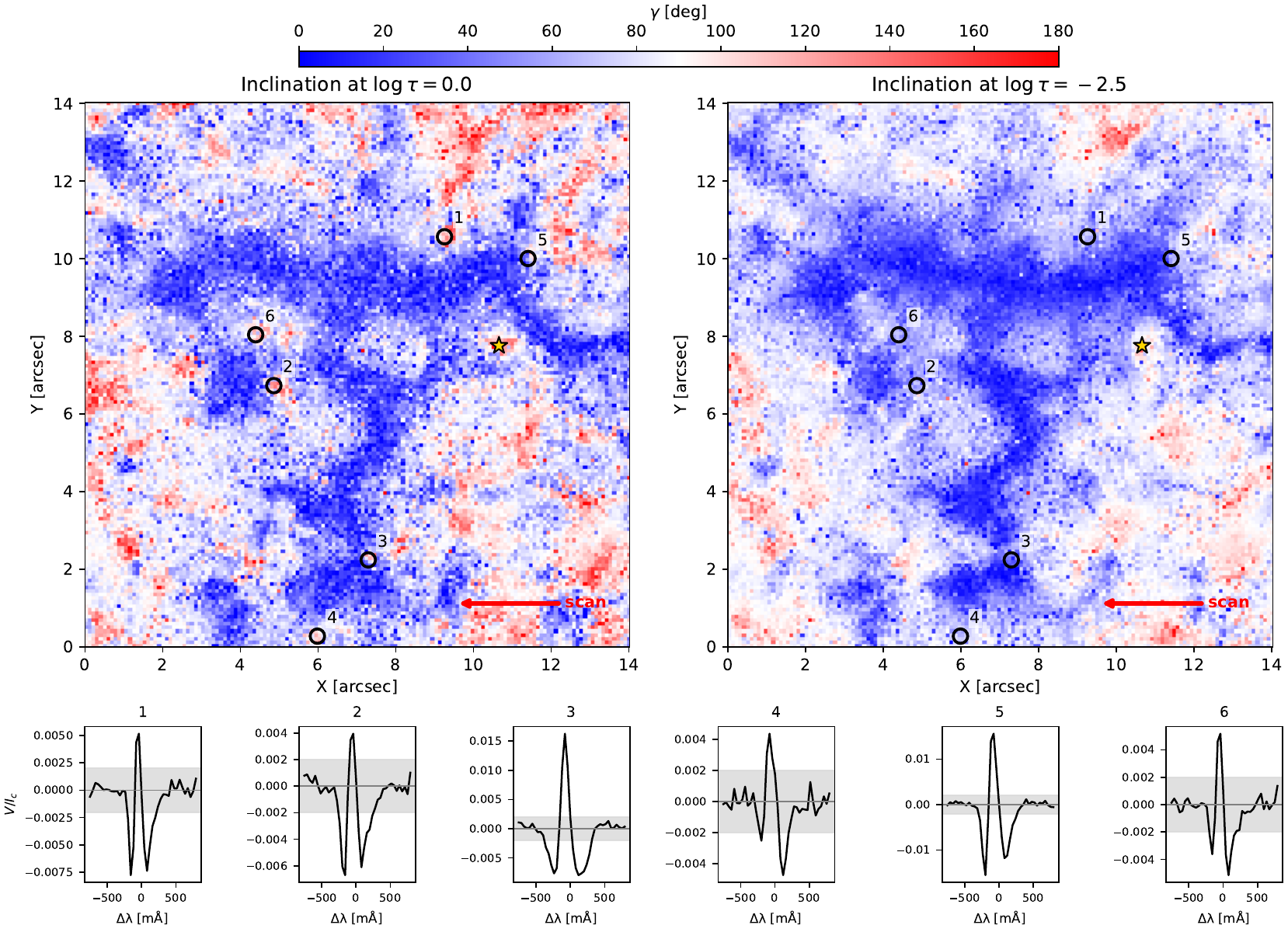}
\caption{Magnetic field inclination in the network region at two representative photospheric depths. \textit{Top left:} Inclination $\gamma$ at $\log \tau = 0$. \textit{Top right:} Inclination at $\log \tau = -2.5$. Black circles mark locations exhibiting three-lobed Stokes $V$ profiles in the Fe~I~$8468$~\AA\ line, where the inferred polarity reverses between the two layers. Labels $1-6$ correspond to the profile examples shown in the lower panels. The gold star marks a distinct location where the observed Stokes profiles show opposite polarity between the photospheric Fe and chromospheric Ca lines, shown in Fig.~\ref{fig:star}. \textit{Bottom panels:} Example Stokes $V$ profiles from the locations marked $1-6$, illustrating the range of three-lobed morphologies observed at the network boundary. The shaded regions show the $5\sigma$ noise level.}
    \label{fig:inclination}
\end{figure*}

Figure~\ref{fig:inclination} reveals the presence of small-scale patches whose magnetic polarity reverses with height. These features appear as localised regions embedded within, or adjacent to, the dominant network field, but exhibit opposite inclination between the two sampled layers. We refer to them here as parasitic patches. We conduct analysis on these with the inversion that did not include the K I lines.

The parasitic patches are preferentially located near the boundaries of the network element. Their spatial distribution is not uniform, but instead confined to specific regions that also exhibit enhanced complexity in the Stokes profiles. As demonstrated by the lower panels in Fig.~\ref{fig:inclination}, these locations are characterised by distinctly three-lobed Stokes $V$ profiles, consistent with the presence of line-of-sight gradients in the magnetic field strength, line-of-sight velocity or inclination \citep{ruedi1992,solanki1993,christoph,campbell2023}. The highlighted pixels are illustrative rather than being presented as an exhaustive list, as there are other parasitic polarity patches that are not highlighted. Further, other network patches in the region scanned by SCIP show these parasitic patches also, as demonstrated in Appendix~\ref{sect:extra_network_examples}.

\begin{figure*}
\includegraphics[width=\linewidth]{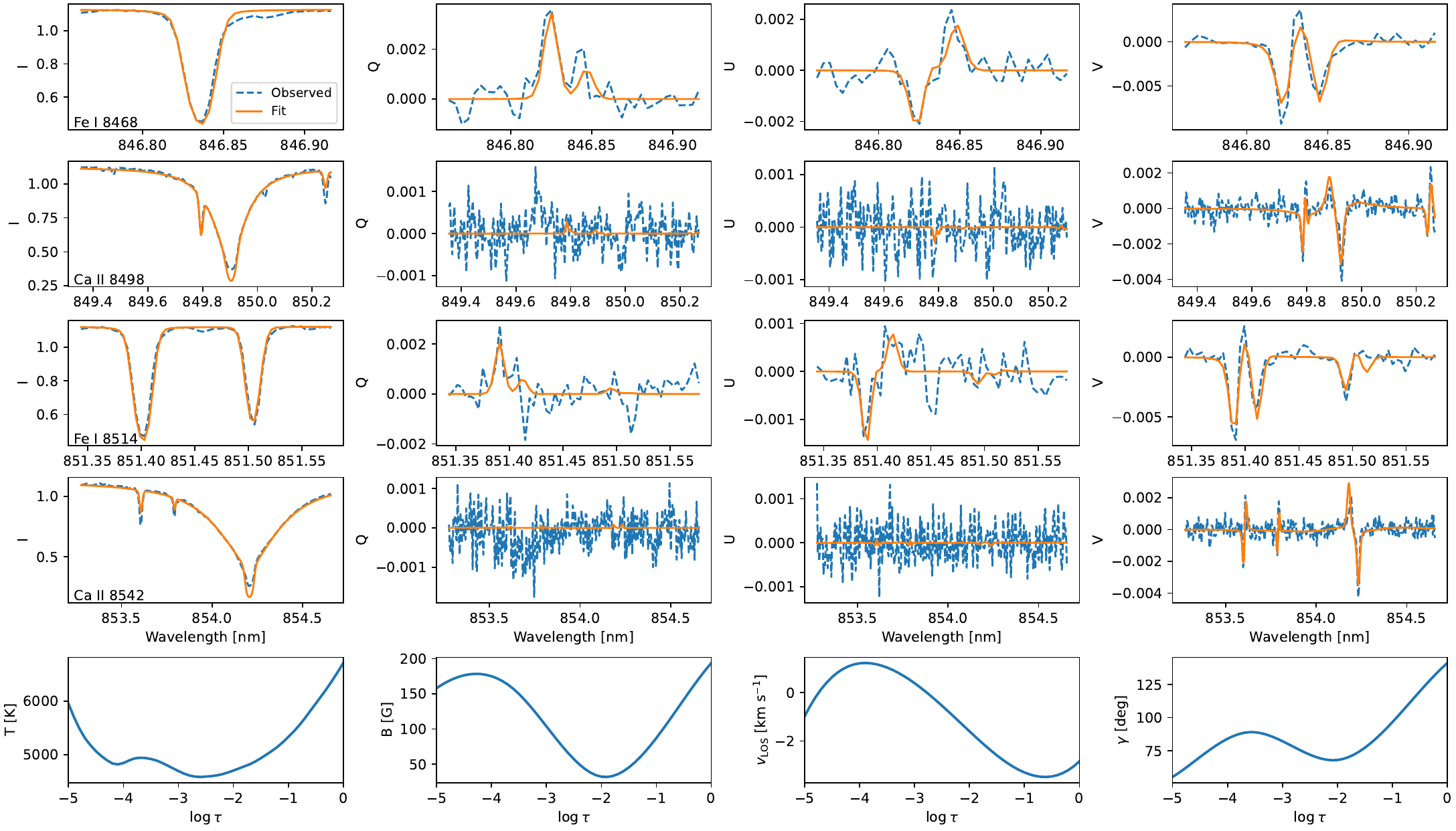}
\caption{Example Stokes profile fits for a boundary pixel identified as exhibiting a three-lobed Stokes $V$ profile in the Fe~I~$8468$~\AA\ line. The inversion reproduces the multi-lobed Stokes $V$ signature in the Fe line, while the Ca lines remain two-lobed in Stokes $V$. The inferred stratification indicates that the observed asymmetry is consistent with strong line-of-sight velocity (and magnetic) gradients.}
\label{fig:3lobedV}
\end{figure*}

Figure~\ref{fig:3lobedV} shows an example inversion result for one of the boundary pixels identified in Fig.~\ref{fig:inclination} as exhibiting a three-lobed Stokes $V$ profile in the Fe\,\textsc{i}~8468~\AA\ line. The observed and fitted Stokes profiles are shown together with the corresponding atmospheric stratification returned by the inversion. The inversion reproduces the asymmetric and multi-lobed Stokes $V$ morphology in the Fe lines while maintaining comparatively simpler chromospheric Stokes $V$ profiles in the Ca\,\textsc{ii} lines. The inferred atmospheric stratification contains strong gradients in the magnetic and velocity parameters along the line of sight, consistent with the interpretation discussed above. This figure is intended only as an illustrative example of the type of inversion solution associated with the parasitic boundary pixels.

\subsection{Polarity reversal between the photosphere and chromosphere}
\begin{figure*}
\includegraphics[width=\linewidth]{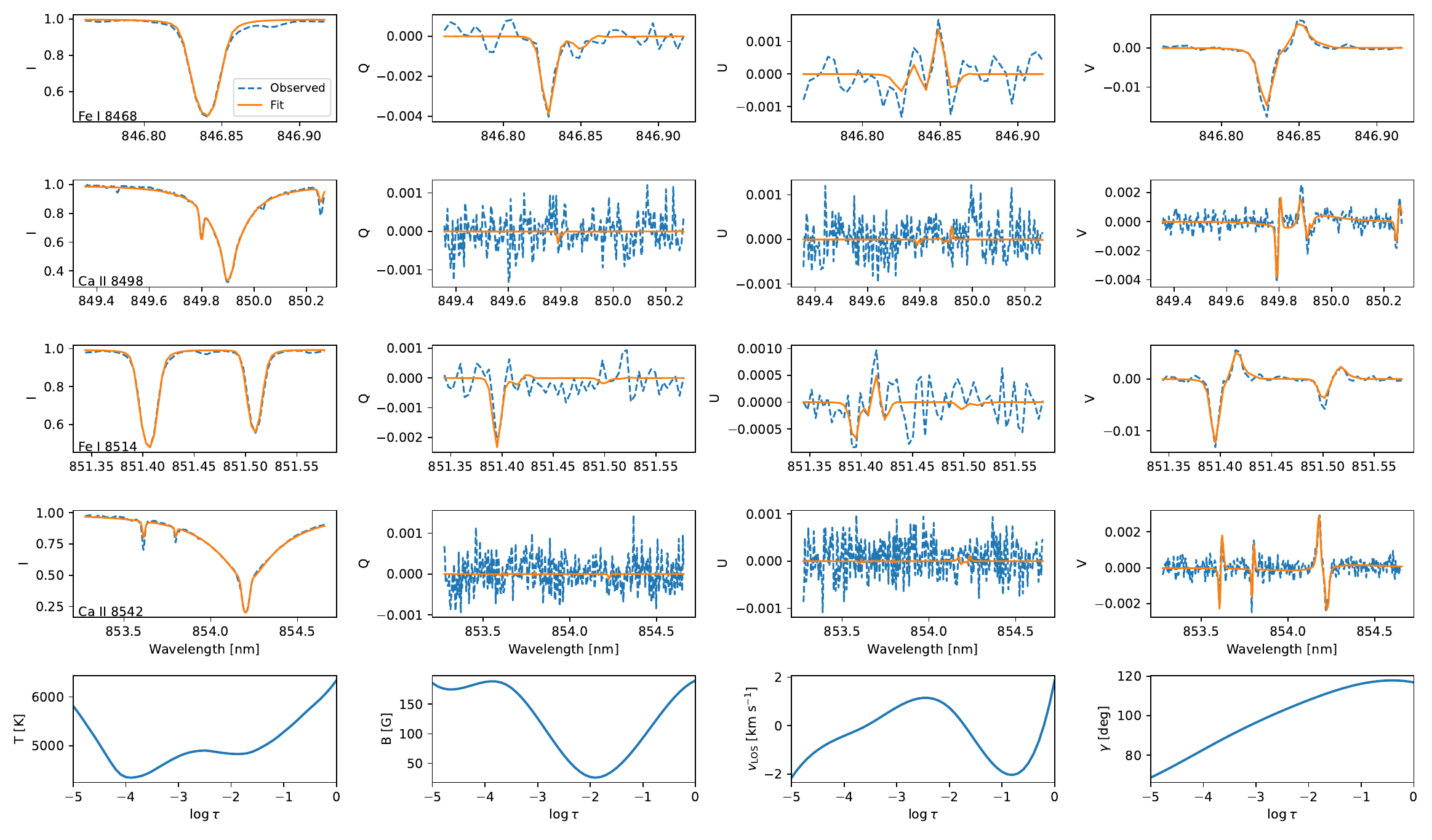}
\caption{Example Stokes profile and inversion fits for the location marked by the star in Fig.~\ref{fig:inclination}. Observed (dashed) and fitted (solid) Stokes profiles are shown for Fe~I~$8468$~\AA, Ca~II~$8498$~\AA, Fe~I~$8514$~\AA and Fe~I~$8515$~\AA, and Ca~II~$8542$~\AA. The Stokes $V$ profiles exhibit opposite polarity between the photospheric Fe lines and the chromospheric Ca lines. The lower panels show the inferred atmospheric stratification as a function of optical depth, including the temperature, magnetic field strength, line-of-sight velocity, magnetic inclination, and magnetic azimuth returned by the inversion.}
\label{fig:star}
\end{figure*}

In the example shown in Figure~\ref{fig:star}, the Stokes $V$ signals from the photospheric Fe\,\textsc{i} and chromospheric Ca\,\textsc{ii} lines exhibit opposite signs. The location of this patch is shown by the star marker in Fig.~\ref{fig:inclination}. This behaviour indicates a significant change in magnetic configuration with height, and is not easily reproduced by a single, depth-independent magnetic field. This could arise as a result of an overlying canopy-like chromospheric field of opposite polarity to the underlying small-scale photospheric magnetic field. The Stokes V profiles are consistent over several pixels. While the profiles shown in Figures~\ref{fig:inclination} and \ref{fig:star} represent individual pixels, they are representative of the types of Stokes $V$ morphologies observed in the regions identified as parasitic in the inclination maps. Taken together, these diagnostics suggest that the parasitic patches are associated with a complex magnetic structure involving vertical gradients.

\section{Discussion}

A significant observational advance of this work is that the canopy-like transverse-field structure is detected through the linear polarisation of a quiet-Sun network element, rather than inferred only from Stokes~$V$ \citep{marian2012}. Previous observations have established that quiet-Sun network flux expands with height, but direct constraints on the resolved azimuthal organisation of the associated transverse field have remained limited by weak linear polarisation signals. The SCIP observations show a coherent boundary ridge of linear polarisation, enhanced $B_\perp$, and a radial azimuthal pattern around a comparatively vertical core. This combination provides a direct magnetic view of the mid-to-upper-photospheric network canopy. The network magnetic field is found to be predominantly vertical in the photosphere, becoming progressively more inclined with height. The inferred morphology is qualitatively consistent with the magnetic expansion observed in larger solar magnetic structures, where comparatively vertical core fields are surrounded by progressively more inclined boundary fields and enhanced transverse magnetic components \citep[e.g.][]{solanki1990,Solanki1999}. 

We also qualitatively compared the inferred magnetic morphology with a quiet-Sun network element from a MURaM simulation. We emphasise that this comparison is intended only as an illustrative physical context rather than a quantitative forward-modelling validation of the observations. Nevertheless, the simulation exhibits qualitatively similar organisation of the transverse magnetic field and lateral azimuthal variation, indicating that the type of structured magnetic morphology inferred from the SCIP observations can arise naturally in realistic magnetoconvective network fields. Perhaps most surprisingly, the MURaM validation further suggests that the azimuthal information encoded in the emergent Fe~I~$8468$~\AA~linear polarisation is weighted toward considerably higher layers than would otherwise be assumed. In the simulation, the minimum median difference between the linear-polarisation azimuth proxy and the underlying magnetic azimuth occurs near $\log\tau\approx-3$. This result does not imply a single formation height for the line, but it suggests that interpretations of observed Fe~I~$8468$~\AA~linear polarisation solely in terms of lower-photospheric magnetic azimuth may overlook a significant contribution from upper-photospheric layers. The result may also help explain why the inclusion of the K~I lines does not substantially alter the inferred azimuthal morphology of the network element, despite their expected sensitivity to higher atmospheric layers \citep{Klines2017}. This conclusion is also substantiated by the response functions shown in Appendix~\ref{sect:RFs}.

In addition, we identify significant small-scale complexity at the boundaries of the network element. Localised, small-scale patches exhibiting three-lobed Stokes $V$ profiles and opposite-polarity line-of-sight fields indicate the presence of strong gradients or unresolved structure within the resolution element. In many cases, the inferred line-of-sight magnetic field changes sign between photospheric depths. This is not unique to this network element, as other network elements in the SCIP field-of-view also show similar parasitic patches (see Appendix~\ref{sect:extra_network_examples}). The presence of localised opposite-polarity features near network boundaries is perhaps related to previous observations of flux emergence and cancellation in quiet-Sun regions, where internetwork fields continuously interact with and modify the network flux \citep{Gosic2014,Gosic2016}. Similar opposite-polarity features were reported by \cite{buhler2015} in Plage regions.

Meanwhile, we find a distinct example which shows opposite polarity between photospheric and chromospheric lines. Apparent polarity reversals between photospheric and chromospheric lines can arise from line formation effects, in particular due to changes in the sign of $\mathrm{d}I/\mathrm{d}\lambda$ in the presence of line core emission \citep{1997A&A...324..763S,gosic2021}. However, opposite-polarity Stokes $V$ signals between Fe\,\textsc{i} and Ca\,\textsc{ii} lines have also been reported and interpreted as signatures of height-dependent magnetic morphology in emerging magnetic structures \citep{2007A&A...469..721S}. In the present case, the absence of emission features in the Ca\,II line cores and the inferred stratification are consistent with a genuine change, rather than an apparent change, in the direction of the magnetic field vector with height. In principle, this represents a possible site for magnetic reconnection. To our knowledge, this is the first detection of such a polarity reversal in a quiet-Sun network element, made possible by the high-resolution, multi-line observations of {\sc Sunrise iii}/SCIP. Photosphere-chromosphere polarity reversals have also been observed by {\sc Sunrise iii} in flares \citep{ishikawa2026,quintero2026}.

\section{Conclusions}

We have analysed high-resolution multi-line spectropolarimetric observations of a quiet-Sun network element obtained with {\sc Sunrise~iii}/SCIP, combining photospheric and chromospheric diagnostics. Our main findings are as follows:

\begin{enumerate}
\item The network magnetic field is strongest and most vertical within the central core of the structure, becoming progressively more inclined with height and distance from the core.
\item The transverse magnetic field is enhanced along the boundaries of the network element, where the magnetic azimuth exhibits a predominantly radial organisation centred on a strong, vertical magnetic core. This geometry is consistent with an expanding magnetic canopy.
\item Localised boundary pixels exhibit complex Stokes $V$ profiles and opposite-polarity line-of-sight fields, indicative of strong gradients or unresolved magnetic structure.
\item A distinct example shows opposite-sign Stokes $V$ profiles between photospheric and chromospheric lines, consistent with a height-dependent reversal of the line-of-sight magnetic field.
\item A representative MURaM network element exhibits a similar radial organisation of the magnetic field, with enhanced transverse field surrounding a comparatively vertical core.
\item We do not detect any azimuthal shear along the line of sight. However, the Fe\,\textsc{i}~8468~\AA\ line is sensitive to the magnetic azimuth much higher in the atmosphere than other commonly observed photospheric spectral lines.
\end{enumerate}

Taken together, these results indicate that the magnetic field within quiet-Sun network elements exhibits significant lateral and vertical structuring, particularly at the boundaries. Such complexity may play an important role in mediating energy transport in the lower solar atmosphere, and underline the importance of combining high spatial and spectral resolution with sensitive polarimetry.


\appendix

\section{Response functions to magnetic azimuth}\label{sect:RFs}

To investigate the atmospheric layers contributing to the observed azimuthal signatures, we computed response functions to perturbations in the magnetic azimuth using the \textsc{DeSIRe} synthesis module. The calculations were performed using a semi-empirical quiet-Sun reference atmosphere with magnetic field strength $B=500$~G, line-of-sight velocity $v_{\rm LOS}=0.5~{\rm km\,s^{-1}}$, magnetic azimuth $\phi=70^\circ$, and magnetic inclination $\gamma=45^\circ$.

\begin{figure*}
\centering
\includegraphics[width=\textwidth]{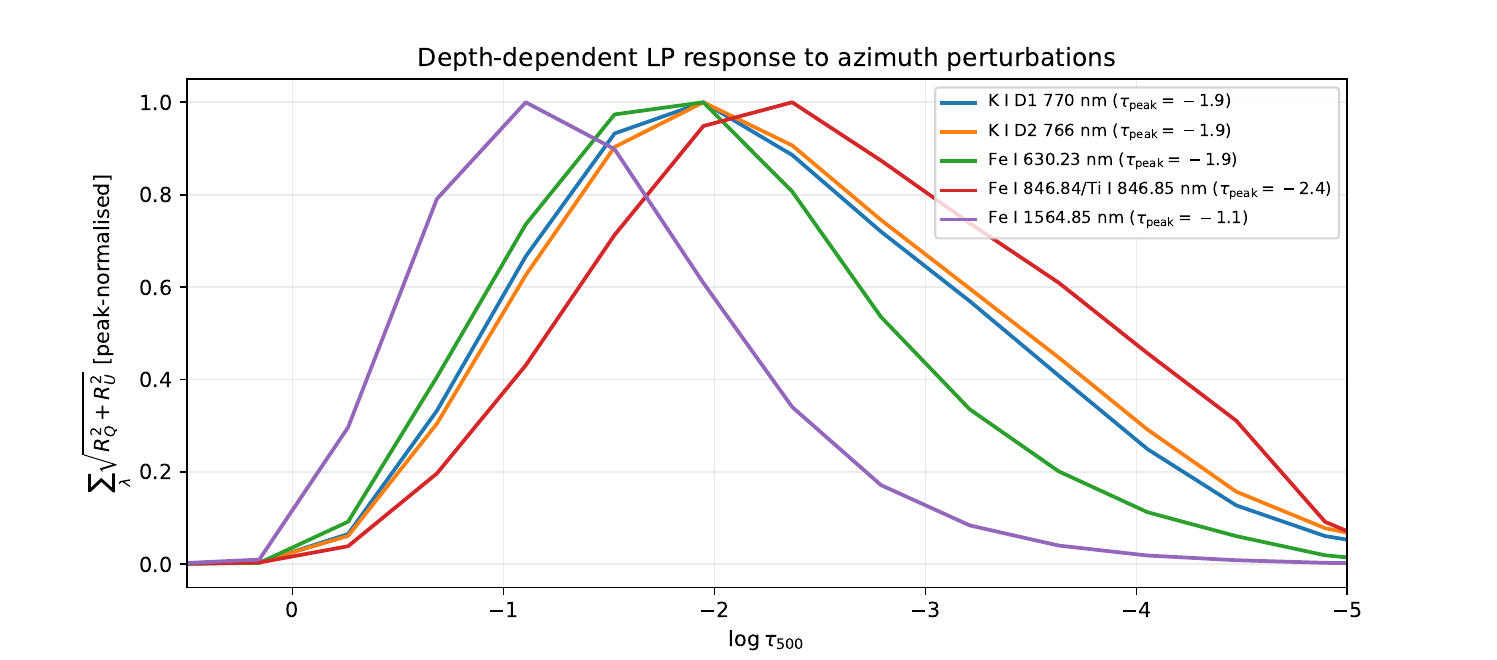}
\caption{Normalised wavelength-integrated response functions to magnetic azimuth for selected spectropolarimetric diagnostics. For each line, the Stokes $Q$ and $U$ response functions were combined and integrated over wavelength before normalisation. The Fe\,\textsc{i}~8468~\AA\ line exhibits substantial sensitivity to upper-photospheric layers, extending to optical depths significantly higher than K\,\textsc{i} D$_1$ and D$_2$.}
\label{fig:RFs}
\end{figure*}

Figure~\ref{fig:RFs} shows the wavelength-integrated response functions to magnetic azimuth for several commonly observed spectropolarimetric diagnostics. For each spectral line, the Stokes $Q$ and $U$ response functions were combined and integrated over wavelength to provide a measure of the total sensitivity of the emergent linear polarisation to perturbations in magnetic azimuth as a function of optical depth. Each curve was subsequently normalised to its peak value to facilitate comparison of the depth dependence.

The response functions demonstrate substantial differences in the atmospheric layers sampled by the various diagnostics. The K\,\textsc{i} D$_1$ and D$_2$ lines do exhibit sensitivity extending into higher layers than the visible Fe\,\textsc{i} lines at 5250~\AA\ and 6302~\AA. However, the Fe\,\textsc{i}~8468~\AA\ line displays a notably extended response toward upper photospheric layers. This behaviour is consistent with the MURaM-based validation presented in Section~\ref{sect:MURaM}, where the azimuth proxy derived from the Fe\,\textsc{i}~8468~\AA\ linear polarisation exhibited its closest correspondence to the underlying magnetic azimuth near $\log\tau\approx-3$.

These calculations do not imply that the Fe\,\textsc{i}~8468~\AA\ line possesses a single formation height, nor that the inferred azimuth proxy can be associated uniquely with a particular optical depth. Nevertheless, they indicate that the linear polarisation observed in this line retains significant sensitivity to upper-photospheric magnetic structure. Consequently, interpretations of Fe\,\textsc{i}~8468~\AA\ azimuth measurements solely in terms of lower-photospheric magnetic fields may overlook an important and perhaps even dominant contribution from higher atmospheric layers.

\section{Parasitic polarity patches in other network elements}\label{sect:extra_network_examples}

\begin{figure*}
    \includegraphics[width=\textwidth]{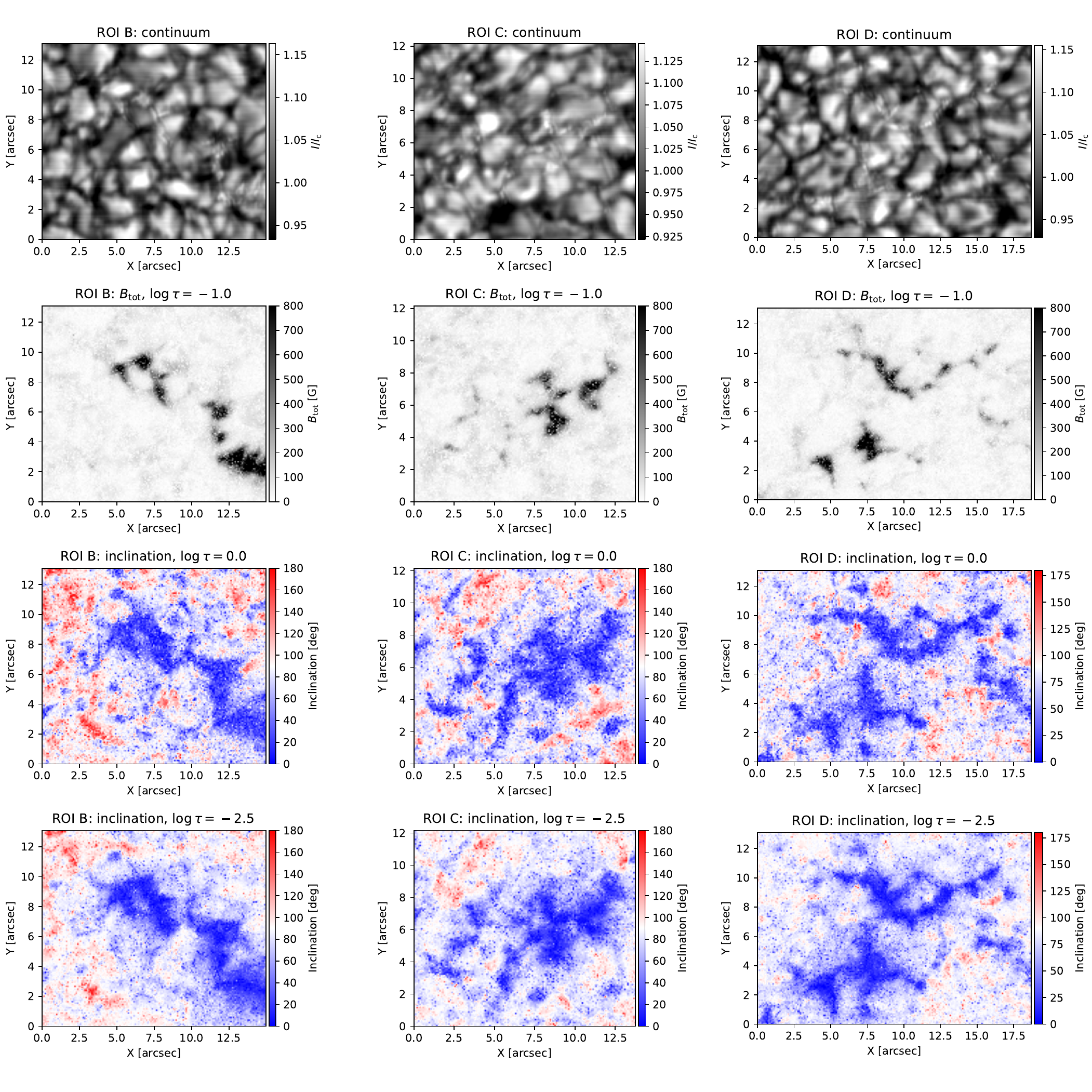}
    \caption{Additional examples of three quiet-Sun network elements identified within the SCIP field of view. The top row shows continuum intensity, the second row shows the inferred magnetic field strength, and the third and fourth rows show the magnetic inclination at $\log\tau=0$ and $\log\tau=-2.5$, respectively. All quantities are derived from the multiline inversions. Similar to the network element analysed in detail throughout this work, each region exhibits a compact magnetic concentration surrounded by smaller-scale patches of differing magnetic inclination.}\label{fig:extra_network_examples}
\end{figure*}

To assess whether the parasitic polarity patches identified in the primary network element are representative, we examined three additional network elements located elsewhere within the SCIP field-of-view, shown in Fig.~\ref{fig:extra_network_examples}. All three additional examples exhibit compact magnetic concentrations accompanied by smaller-scale patches of contrasting magnetic inclination in their immediate surroundings. The opposite polarity parasitic patches are visible at $\log\tau=0$ and mostly vanish in the upper-photosphere (as shown by the $\log\tau=-2.5$ map). These examples suggest that the mixed-polarity patches described above are not restricted to a single network element, but are present in multiple magnetic concentrations across the observed quiet-Sun field. In Fig.~\ref{fig:extra_network_examples} the three additional network elements are labelled as regions of interest (ROI) B, C, and D.

\begin{acknowledgments}

RJC, MM and DBJ acknowledge support from the Science
and Technology Facilities Council (STFC) under grant
No.ST/X000923/1. Sunrise III is supported by funding from the Max-Planck-Förderstiftung (Max Planck Foundation), NASA under Grants \#80NSSC18K0934 and \#80NSSC24M0024 (“Heliophysics Low Cost Access to Space” program), and the ISAS/JAXA Small Mission-of-Opportunity program and JSPS KAKENHI Grant Numbers JP18H05234 and JP23K25916. This research has received financial support from the European Union’s Horizon 2020 research and innovation program under grant agreement No. 824135 (SOLARNET) and No. 101097844 (WINSUN) from the European Research Council (ERC). It has also been funded by the Deutsches Zentrum für Luft- und Raumfahrt e.V. (DLR, grant no. 50 OO 1608). The Spanish contributions have been funded by the Spanish MCIN/AEI under projects RTI2018-096886-B-C5, PID2021-125325OB-C5, and PID2024-156066OB-C5, and from “Center of Excellence Severo Ochoa” awards to IAA-CSIC (SEV-2017-0709, CEX2021-001131-S), all co-funded by European REDEF funds, “A way of making Europe".

\end{acknowledgments}

Data availability statement: The Sunrise observations are openly available via the Sunrise-3 Mission Archive (https://sr3data.mps.mpg.de).

\bibliography{sample701.bib}
\bibliographystyle{aasjournalv7}

\end{document}